An Entropic Story
Jeremy Bernstein
Stevens Institute of Technology

Abstract: This is a pedagogical survey of the concept of entropyy from Clausius to black holes.

"We might call *S* the *transformation content* of the body, just as we termed the magnitude *U* its *thermal and ergonal content*. But as I hold it to be better terms for important magnitudes from the ancient languages, so that they may be adopted unchanged in all modern languages, I propose to call the magnitude S the entropy of the body, from the Greek word ôńďđç, *transformation*. I have intentionally formed the word *entropy* so as to be as similar as possible to the word *energy*; for the two magnitudes to be denoted by these words are so nearly allied their physical meanings, that a certain similarity in designation appears to be desirable." Rudolf Clausius-1865

" When I put my hot tea next to a cool tea they come to a common temperature. I have committed a crime. I have increased the entropy of the universe. Jakob Bekenstein came by and I said to him if a black hole comes by and I drop the cups behind it I can conceal my crime. He was upset by this and came back in a few months to tell me that black holes have entropy."
John Wheeler, The Web of Stories

"The remarks on *negative entropy* have met with doubt and opposition from physicist colleagues. Let me say first, that if I had been catering to them alone I should have let the discussion turn on *free energy* instead. It is the more familiar notion in this context. But this highly technical term seemed linguistically too near *energy* for making the average reader alive to the contrast between the two things."
Erwin Schrödinger, What is Life, Canto, Cambridge, 1967,p 74.
"Nobody really knows what entropy really is." John von Neumann[1]

I first heard of the concept of entropy in 1949 when as a somewhat retarded sophomore I took Freshman physics. My teacher was Wendell Furry, a nice man and a very competent physicist who a few years later got caught up in the McCarthy hearings. Furry had had a brief flirtation with Communism-a surprise to me-and was the subject of a hearing in Boston which I attended. Harvard did not fire him but insisted that he testify fully about his activities. He refused to name names and was cited for contempt, a case that was later dropped. None of this was in any way evident in his lectures and indeed he cultivated the appearance of a sort of mid-western hick.

Somewhere in the spring of 1949 he introduced us to thermodynamics and the concept of entropy. From the beginning I simply could not get my head around it.

In the first place there were the units-ergs per degree Kelvin. They made no sense to me. On the other hand ergs per degree Kelvin per kilogram made perfect sense. This at least was the specific heat of something and indeed we did experiments to measure the specific heat of something or other. But no experiment could tell us its entropy. None the less a change in entropy was well-defined. There was always, we were told, a change of entropy. But why did this matter? Why bother? I don't mean to appear frivolous but I just couldn't make sense of it.

I think that I would have been helped by an historical introduction but this was not what interested Furry. He was a meat and potatoes type-solve the problems in the book and on the exams. I was lucky to get a B in the course. On the other hand energy made sense to me. That could be measured and was conserved. My first great physics teacher Philipp Frank however gave me something to think about. When he was a young man he wrote a paper on the conservation of energy. He argued that it was a tautology which could never be disproved since whenever it seemed to be violated you just added on another form of energy which had just happened with radioactive decays. He got a letter from Einstein-the beginning of a life-long friendship. Einstein said he was right but missed the point. The point was that you had to add on very few of these new energies; $E=mc^2$ being the recent example. It was this economy that gave the energy conservation its value. I wished I had asked Professor Frank about entropy. He was after all a student of Boltzmann.

It is quite evident that the development of thermodynamics was related to the use of steam as a motive power. I have often thought that the advent of relativity had to do with the electrification of cities. Einstein's father was in the electricity business. The high point of the early work on thermodynamics came with the publication in 1824 of Sadi Carnot's memoire *Réflexions sur la Puissance Motrice du Feu.*[2] All of the reasoning is done in terms of the "caloric" theory of heat. Heat was transmitted by the exchange of a substance which was called "caloric." In some sense heat was caloric. Early in the memoire Carnot describes the cycle that bears him name. One wants to create a certain *puissance motrice*-motor power-by a series of steps that are reversible. For example the first step is to use the caloric in a body A to make steam. The steam is used to elevate a piston and at the end of the cycle the steam is condensed into water. He notes that if it was possible to find a cycle that produced more motor power than this one could use the excess to run his cycle in reverse and recommence the operation producing a perpetual motion machine. This he says contradicts the laws of *"la saine physique"* – "sound physics." [3] If one had asked him why, I . wonder what he would have said. In a footnote he remarks that, " If it was possible, there would be no point in searching in water or air courants, or in the combustibles for this motor power; we would have at our disposition a source at our disposal that was inexhaustible." [4]

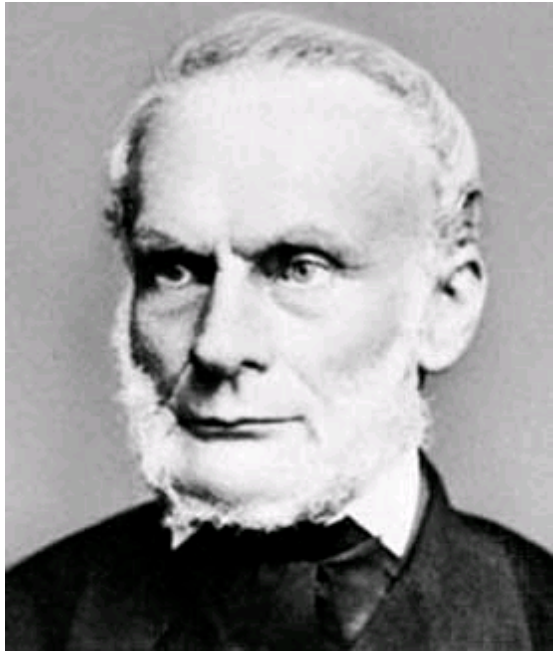

Rudolf Clausius

It is to the German physicist Rudolf Clausius that we owe the concept of entropy. He was born in Köslin in 1822. He was one of a large number of siblings and received his primary education in a school run by his father. He attended the University of Berlin where he studied physics and mathematics later taking his PhD from the University of Halle. He had some appointments in Germany but in 1855 he became a professor of physics in the Swiss Polytechnic Institute in Zurich-Einstein's "poly" from which he graduated in 1900. When Clausius left in 1867 for a post in Germany he was not replaced and even in Einstein's day there was no professor of theoretical physics.

Einstein learned most of his theoretical physics on his own. The one thing he learned to love from the beginning was thermodynamics. One wonders if he studied Clausius's papers.

Clausius was not only a profound physicist but he was prolific Between 1850 and 1865 he published nine substantial memoirs on thermodynamics. In 1875 he published a book length memoir whose English title is "The Mechanical Theory of Heat." The English translation was published in 1879 and is available in the form of a book published by Biblio Bazar.[5] In his preface Clausius tells us that this is a revised edition of his earlier memoir and represents his latest thoughts. In addition in 1857 he published a paper called "Über die Art der Bewegung wir Warme nennen"-"On the nature of the motion we call heat." It was published in English under this title in the *Philosophical Magazine.*[6] In this paper, which I will now discuss, he founds kinetic theory.

Clausius tells us that he was inspired to publish this paper sooner that he had expected to because of some similar work by the German physicist August Krönig. Clausius notes that Krönig had only considered motions of the "molecules" in straight lines and not possible molecular degrees of freedom such as rotation and vibration. What struck me is that both of these physicists accept as a matter of fact the atomic structure of matter such as gasses. I bring this up because towards the end of the century this view was challenged especially by the physicist-philosopher Ernst Mach. He simply would not accept the "reality" of atoms. In a debate with Boltzman he asked, "Have you seen one?" This question left Boltzmann stupefied. Also gone is the caloric theory of heat. In his 1850 paper published also in English under the title "On the Moving Force of Heat and the Laws regarding the Nature of Heat itself which are deducible therefrom"[7] there is the following brief dismissal of caloric,

"If we assume that heat, like matter, cannot be lessened in quantity [ as would be the case if heat were, a material object like caloric] we must also assume that it cannot be increased; but it is almost impossible to explain the ascension of temperature by friction otherwise than by assuming an actual increase of heat." This was written some twenty five years after the publication of Carnot's memoir. The caloric is gone.

In Clausius's paper he gives a derivation of the relation between the temperature and the average kinetic energy of the molecules making up an ideal gas which you could teach in a present day class. The only change you would have to make is to replace the term *vis viva*-"living force"- by "kinetic energy." Clausius begins by calculating the pressure of the walls of a container which he takes for simplicity to be parallel planes. He assumes that the molecules collide with the wall elastically-the velocity is unchanged in the collision and the angle of incidence is equal to the angle of reflection. He then finds the number of collisions per second with the wall for all collisions in which there is a given angle of incidence of molecules that are moving with an average speed that he calls 'u.' He calls 'n' the total number of molecules and assuming that their density is uniform he finds the number in an angular interval. He can now integrate over angles to find what he calls the "motion" imparted to a unit area of the surface. This leads him to an expression for the pressure on the wall, using his notation

$$p = nmu^2/3áh \qquad (1)$$

where á is the area of the wall, h is the distance between the wall and m is the mass of a molecule. The quantity áh is of course the volume of the container v. Using the ideal gas law

$$Pv = const.T \qquad (2)$$

he arrives at

$$n\,mu^2/2 = const.T, \qquad (3)$$

with a different constant.

Using experimental data he finds the average speed of the molecules in a gas like oxygen at the temperature of melting ice. He finds 461 meters per second for this which is in general agreement with modern values but this was done in 1857! There is no discussion in this paper of either entropy or the second law. For that we turn to his book.

As I was reading his book two questions occurred to me. How did he chose the letter 'S' for 'entropy and how did he arrive at the definition S=Q/T where Q is the heat and T the temperature? He does not address the first question but it has been suggested that he used the initial of Carnot's first name 'Sadi.' The letter 'C' was already taken for specific heat. On the second question we can follow Clausius's logic. He considers a reversible process in which a certain amount of heat $Q_1$ is transferred from a reservoir and in this process an amount $Q_2$ is received by another reservoir, then he argues that the ratio $Q_1/Q_2$ can only be a function of the temperatures $T_1$ and $T_2$; ie,

$$Q_1/Q_2 = \ddot{O}(T_1, T_2). \tag{4}$$

The question then becomes the determination of Ö. To this end he uses the equation of state

$$pV = RT \tag{5}$$

to study the Carnot cycle. He defines R as the difference between the specific heats $C_p$ and $C_V$; ie, $R = C_p - C_V$. He does not discuss specfic heat per mole but rather talks about "mechanical units." I will follow his derivation but change the notation to make it simpler. He makes use of what must have been a fairly recent formulation of the Carnot cycle given by the French engineer physicist Benoit Paul Emile Clapeyron. Clapeyron introduced the now familiar diagram

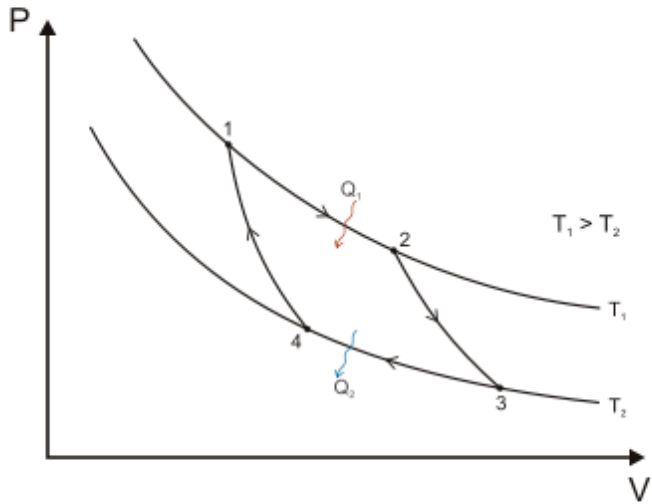

Fig.1 The Clapeyron diagram for the Carnot cycle.

The two upper and lower hyperbolas represent the isothermal expansions and contractions of the gas. The change of heat is determined by the equation

$$dQ = RTdV/V. \tag{6}$$

or

$$Q_u = RT_1 \log(V_2/V_1) \tag{7}$$

for the upper branch and

$$Q_l = RT_2 \log(V_3/V_4) \tag{8}$$

for the lower. The side branches of the curve represent and adiabatic expansion and contraction of the gas with dQ=0. Thus during this portion of the cycle we have

$$dT/T = -(C_p-C_v)/C_p dV /V \equiv (k-1)dV/V. \qquad (9)$$

Here $k=C_p/C_V$ and should not be confused with the k that is often used for the Boltzmann constant.

Between any two temperatures and volumes we have

$$T/T' = (V'/V)^{(k-1)} \qquad (10)$$

Or using the notation of the graph

$$V_1/V_4 = V_2/V_3 \qquad (11)$$

Or

$$V_2/V_1 = V_3/V_4 \qquad (12)$$

Hence we have

$$Q_1/Q_2 = T_1/T_2 \qquad (13)$$

Or

$$Q_1/T_1 = Q_2/T_2 \qquad (14)$$

Originally Clausius had called this relation "the equivalence of transformations" but in 1865 he introduced the term "entropy" and this equation was simply the equivalence of entropy in these two cycles or the statement in this reversible cycle that the entropy was conserved. He generalized this to an arbitrary reversible cycle with the equation

$$dS = \int dQ/T = 0 \qquad (15)$$

where the integral is around the cycle. Clausius notes that in general

$$S = S_0 + \int dQ/T \qquad (16)$$

so that his entropy cannot be determined without specifying some initial entropy. This, as I have mentioned, is one of the things that troubled me as an undergraduate. What was the entropy of something? In making this relation Clausius notes that while dQ is not a true differential, dS is. He does not use this language but he shows that the temperature is an integrating factor making the entropy a function of state.

In a later chapter of his book Clausius discusses what he calls the "second main principle". The first being the conservation of energy. Now he considers irreversible processes and states,

"For one kind of transformation, viz. the passage of heat between bodies at different temperatures, it was taken as a fundamental principal depending on the nature of heat, that the passage from a lower to a higher temperature, which represents negative transformation cannot take place without compensation. On this rested the proof that the sum of all the transformations in a cyclical process could not be negative, because if any negative transformation remained over at the end, it could always be reduced to the case of a passage from a lower to a higher temperature."[8]

In short for all such processes

$$\Delta S \geq 0. \quad\quad\quad\quad (17)$$

This led to his famous couplet

*Die energie der Welt ist constant*
*die Entropie strebt einen Maximum zu .*

The energy of the world is constant
The entropy strives for a maximum.

The reader will have noticed that there is not a word in this discussion about probabilities. This is not surprising. Clausius believed and continued to believe that the second law had the same significance as the first. The increase in entropy was a fact and not a probable outcome. It has been commented that as far as one knows Clausius never commented on Boltzmann's probabilistic definition of entropy which he offered beginning in 1877. The two men were certainly familiar with each others work. They even had discussions on some priority issues but on Boltzmann's definition of entropy-shown below-there was on Clausius's part silence.

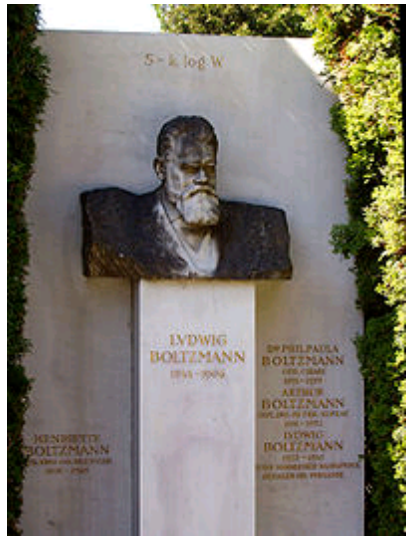

Fig.2 Boltzmann's grave.

The photo is of Boltzmann's grave in Vienna with the entropy formula S=klogW above the bust. 'k' is of course the Boltzmann constant k=1.380 6488(13)×10$^{-16}$ ergs per Kelvin-the dimension of entropy.  Boltzmann died by his own hand in 1906 while Clausius died in 1888.

I will return to the statistical mechanics of entropy after I make a digression into the Third Law of Thermodynamics which was first suggested by the physical chemist Walther Nernst in 1906. Nernst was born in West Prussia in 1864. His life overlapped with that of Clausius and he also studied for a time with Boltzmann. His father was a judge. In 1892 he married one Emma Lohmeyer who bore him five children; two sons and three daughters. His two sons were killed fighting in the First World War in which Nernst did some operational research. Two of his daughters married Jews. With the rise of the Nazis Nernst became an outspoken opponent of the regime and resigned from his official positions. Nernst, who had visited Einstein in Zurich in 1910 and later became his colleague in Berlin was one of the very few scientists who remained in Germany in the war for whom Einstein had any use. Max von Laue was another. Nernst won the Nobel Prize for chemistry in 1920 and died in Germany in 1941.

The usual statement of Nernst's third law is not due to Nernst but rather Planck. Planck's version is that the entropy of a system tends to zero when its absolute temperature tends to zero. Nernst did not regard the notion of entropy as being

"intelligible" as opposed to free energy which was measurable. Of course Nernst knew the relation between the free energy and the entropy. If U is the internal energy then the Helmholtz free energy for available work A-

'A' stands for 'arbeit"-work- is defined to be

$$A=U-TS. \qquad (18)$$

To get at Nernst's version of the second law consider a transformation in which one state is transformed into another only by the absorption of heat. No work is done. Thus

$$dA=dU-SdT-TdS \qquad (19)$$

As

$$dU=dQ \qquad (20)$$

and

$$dS=dQ/T \qquad (21)$$

and

$$dA/dT=-S=(A-U)/T \qquad (22)$$

or

$$TdA/dT=A-U. \qquad (23)$$

What Nernst claimed was that dA/dT went to zero as T went to zero. From this he argued that at T=0, A=U and that dU/dT also went to zero which implied that specific heats went to zero something that seemed to be borne out by experiment. As far as entropy is concerned Nernst claimed only that its change went to zero as the temperature went to zero. It was Planck who claimed that the entropy itself went to zero. It is now known that this law has some exceptions, At the end of this essay I will discuss whether black hole entropy is among them. Now to Boltzmann.

It would appear as if at first Boltzmann agreed with Clausius that the second law was as absolute as the first. But from the 1860's onward probability entered his kinetic theory considerations. Finally in 1877 he presented his definition of entropy. I give here the translation of the relevant paragraph.[9]

"The initial state will essentially be a very improbable one. The system will run from this state to ever more probable states until it reaches the most probable state, which is the state of thermal equilibrium. If we apply this to the second law, we can identify the quantity that is usually called entropy with the probability of that state."

That Boltzmann meant the logarithm of this probability is made clear later in the paper. Interestingly he did not use 'W'-*wahrscheinlichkeit*"-"probablilty" for this. It was supplied by Planck. The 'W' on Boltzmann's tomb cannot stand for probability Planck or no Planck.. In any event I recall how pleased I was with this explanation of the second law. I had always found the thermodynamics explanation, if that is what it was, as circular and incomprehensible. If one could have reactions in which entropy decreased then one could imagine a situation in which if a reservoir of heat A could deposit spontaneously heat in a reservoir B , also the inverse could be possible. Putting these two operations together one could produce a perpetual motion machine. But if one asked why this was impossible one was told that it violated the second law-a perfectly circular argument. But here Boltzmann was proposing a reason for the second law. Systems went from less probable to more probable states-what could be more plausible-then finally reaching an equilibrium state-stasis-which maximized the entropy. It seemed to fit together like a glove. However, I soon learned that it was not so simple. First there was Loschmidt's "paradox."

Joseph Loschmidt was born in what was then the Austrian empire in 1821. He was what one would now call a physical chemist. He invented a notation for molecules which is close to the one in present use and he used the kinetic theory to estimate the size of molecules in air. He wanted to determine the number in say a cubic centimeter , a number that now bears his name. In 1868 he became a faculty member at the University of Vienna, Boltzmann joined the faculty in 1873 but three years later he returned to the University of Graz where he spent fourteen years until he returned to Vienna. It was here that Professor Frank heard him lecture. Professor Frank, who later succeeded Einstein in Prague, and had heard them all, once told me that Boltzmann was the most brilliant lecturer he had ever heard.
Loschmidt was in an excellent position to follow Boltzmann's early work on entropy and in 1876 he presented what is often referred to as his "paradox."

I find the use of "paradox" in this context to be a misnomer. To me a paradox is a choice of opposites both of which lead to a self-contradiction. The example I always keep in mind is Bertrand Russell's barber's paradox. A barber tries to decide whether or not he belongs to the class of barbers who shave all and only men who do not shave themselves. Loschmidt's observation is nothing like this. Recall that Boltzmann ascribed the increase in entropy of say a gas to the evolution of a given configuration described by the positions and momenta of the gas molecules at some given time to a new and more probable configuration. We know intuitively what this means. If for example all the molecules are initially confined to the corner of a box, an improbable configuration, they will spread out to fill the box. But the mechanism by which this happens is molecular collisions. Loschmidt made the simple observation that if some sequence of collisions led to a more probable configuration and hence an increase In entropy then reversing all the momenta involved would lead to just the opposite. This reversal was allowed because the underlying dynamics was time reversible. I do not call this a "paradox". It seems to me more like a statement of fact. Whatever you call it, volumes have been written. It seems to me that the correct conclusion is straightforward. This reversal can happen in principle but re-creating the initial conditions has a vanishingly small probability. In short, the second law, unlike the first, is a statement about probable behavior. Pace Clausius. In somewhat the same category is the Poincaré "recurrence theorem," published in 1890. Loosely speaking it says that given some initial condition involving the positions and momenta of all the molecules, if one waits long enough, the system will approach a configuration arbitrarily close to this one. "Long enough," might be the age of the universe. For this reason Boltzmann did not take this very seriously.

Up to this point I have been using the notion of probability rather cavalierly. Probability simply speaking involves counting. If I have an urn containing black and white balls and I want to know the probability of drawing a black ball on my next try I must count the total number of balls and the number that are black. But the classical statistical theory is not like this. I have a space which contains a continuum of points each one specified by six coordinates-three for position and three for momentum. If there are N molecules I would like to know the probability for some configuration of these points. How do I do the counting? This was the problem that faced Boltzmann. His device was to divide the "configuration space" into cubes. Given a point in phase space specified symbolically by p-momentum-and q position, then the points in the box with the side $\Delta p \Delta q$ will be within these errors of p and q. This box has volume $(\Delta p \Delta q)^3$. The number of points, each one corresponding to a state of the system, will be proportional to this volume. But classically Δp and Δq are arbitrary and so is this volume. But in the expression for the entropy it is the logarithm that enters. Thus there is an arbitrary additive constant that depends on the choice of volume. This is as it was in the thermodynamic case where there was also an arbitrary constant depending on the initial entropy. In a change in entropy this constant cancels.

Once quantum mechanics was discovered it was realized that it was ΔpΔq~ℏ, Planck's constant which sets the scale for the size of these boxes in phase space. One cannot shrink the volume below the uncertainty principle limit. But of course this is not the only consequence of the quantization. The observables are quantized. To take an example if we consider a cubic box with a side L, then the ith energy level is given by

$$E_i = \hbar^2 \pi^2 / 2mL^2 (n_{i1}^2 + n_{i2}^2 + n_{i3}^2) \qquad (24)$$

The n's here are integers that can range upwards from one. If there are N particles then the number of degrees of freedom f is given by 3N. There are two interesting limits. The "classical limit" is when the n's become very large. The energy for a given state E is gotten by summing over the $E_i$'s and the number of states is proportional in this limit to $E^f$ so the entropy is proportional to f log(E). At the opposite end is the entropy of the ground state. This is the state in which all the $n_i$'s are one. This state is non-degenerate so the number of such states is one. As log(1)=0 this is a state of zero entropy. There are quantum ground states involving for example spin which are degenerate. For these states the entropy is not zero. If these states can be realized in the limit of vanishing temperature then we have a violation of Plabck's version of the third law but not Nernst's .

These considerations really involve "old quantum mechanics." In 1927 von Neumann published a paper in which the then new quantum mechanics entered.[10] In 1932 he expanded these ideas in his great book *Matematische Grundlagen der Quantenmechanik*[11] . His definition of entropy is closer to that of the great American theorist Willard Gibbs, a contemporary of Bolzmann, than it is to that of Boltzmann. Like Boltzmann, Gibbs divided the phase space into boxes. But to each of these boxes was associated a probability, $p_i$, that system points would be located therein. The Gibbs entropy is by definition

$$S = -k \sum_i p_i \log(p_i), \qquad (25)$$

where the $p_i$ are positive and

$$\sum_i p_i = 1. \qquad (26)$$

This means that the logs are negative which accounts for the minus sign. In the case in which the probability per microstate is simply one over the number of microstates and hence the same for all bins this reduces to the Boltzmann entropy.

What corresponds to a classical ensemble in quantum mechanics is a state that can be decomposed into a sum of states consisting of orthonormal eigen functions of some observable.[12] Or

$$\Psi(q,t) = \sum_i a(i,t) u(i,q). \qquad (27)$$

Here the u's are the eigenfunctions and the a's are complex probability amplitudes. The q's are a shorthand for all the coordinates. We may use the a's to form the "density matrix" –$\rho$–for this ensemble. Thus if the total number of such states is N [13]

$$\rho_{nm} = 1/N \sum_{i=1}^{N} a_i^*(m,t) a(n,t)_i . \qquad (28)$$

Written this way the probability of finding the system in the nth state chosen at random from the ensemble is given by $\rho_{nn}$. For a pure state only one term in the sum survives and $\rho_{nm} = a^*_m a_n$. In this case $\rho^2 = \rho$ so that $\rho$ is a projection operator with eigenvalues zero or one.

The von Neumann entropy is defined to be

$$S = -k \operatorname{Tr}(\rho \log(\rho)). \qquad (29)$$

Here Tr stands for the trace which is in any representation of the matrix the sum of the diagonal elements. Since $\rho$ is a Hermitian matrix it can be diagonalized by a unitary trans formation of the basis. The diagonal elements in this basis are the eigen values which are real. If we call these values $p_i$ then the von Neumann entropy can be written as

$$S = -k \sum_i p_i \log(p_i). \qquad (30)$$

This has the same form as the Gibbs entropy although the underlying physics is different. Note that if you have a pure state all the eigen values are zero except one of them which is one. The entropy is zero. The von Neumann entropy is a measure of what is often called "entanglement." A pure state is unentangled. The von Neumann entropy has a number of wonderful properties which one is tempted to linger over. But I want to end this rather lengthy *tour d'horizon* with a brief excursion into the entropy of black holes.

A number of years ago I interviewed John Wheeler for a profile which I called "Retarded Learner"[14] This was the way that Wheeler referred to himself. During this interview he told me about Bekenstein and the coffee-the anecdote I quoted above. I am embarrassed to say that I did not understand what he was talking about. I wish I had because there are questions I would have very much liked to ask. One of them was what Wheeler understood when he asked his question. He was given to asking students questions that he knew or guessed the answer to, to give them the satisfaction of making the discovery, He did this with Feynman. I believe that he understood, and was trying to convey this to Bekenstein, that classical black holes appeared to violate the second law of thermodynamics. Why?

When a classical black hole is formed as a massive star collapses gravitationally an "event horizon" is formed. Seen from the outside the radiation from the hole is more and more red-shifted until it essentially stops being emitted. Once this happens the classical black hole is black. It acts as a perfect absorber. This means that its temperature vanishes as viewed from the outside. The entropy diverges. On the inside there is a raging inferno. Suppose you drop something into the hole. This is a loss of entropy for you but nothing compensates it. Thus you have a violation of the second law. This is what I think that Wheeler was telling Bekenstein. In his 1972 Princeton PhD thesis Bekenstein proposed that the solution to this dilemma was that a quantum mechanical black hole did have a finite entropy. He produced a formula for it in which the entropy was proportional to the surface area of the hole and inversely proportional to Planck's constant. Curiously he did not discuss what this implied about the temperature. This was left in 1974 to Stephen Hawking. What Hawking showed is that the simplest kind of black hole-the so-called Schwarzschild black hole- emits black body radiation which to an external observer would appear to have a temperature of

$$T = \frac{\hbar c^3}{8\pi G M k_b} \quad (\approx \frac{1.227 \times 10^{23} \, kg}{M} K) \quad (31)$$

Here M is the mass of the black hole. The mass of the Sun is about $2 \times 10^{30}$ kg so for a typical black hole the temperature is very small. On the other hand using the fact that we are dealing with black body radiation one can argue that the entropy S is given in this case by

$$S = kc^3 A / 4G\hbar \quad (32)$$

where A is the surface area of the black hole given in this case by

$$A = 16\pi (GM/c^2)^2. \quad (33)$$

If we put these things together we see that the entropy in this case goes as $1/T^2$. Hence we have a violation of Nernst's theorem big time. The Schwarzschild black hole is rather special. It has no charge and does not rotate. Charge, mass and angular momentum is what characterizes a black hole to an external observer. This changes the entropy which goes to a finite value as the temperature goes to zero-still a violation of the third law.[15] There is much discussion in the literature about this and I think that its significance is a matter of debate. Finally there is the very interesting subject of "duality." Some string theories can be re-written so that they become the dynamics of black holes. One can do statistical mechanics of these strings and it is claimed that making the duality transformation one can derive the Hawking temperature.[16] These are deep waters.

My goal has been to produce a kind of tour guide to this subject. If I have succeeded then the reader may be encouraged to make longer visits.

Appendix: More About Black Holes

In 1949, just as I was struggling to understand entropy, the book The Mathematical Theory of Communication by Claude Shannon and Warren Weaver appeared.[17] At this time Shannon was at the Bell Labs and the year before had published his seminal paper on the subject. It will be recalled that in the book, and the paper, Shannon introduced what he called "entropy." Actually he had not known what to call it and asked von Neumann. Von Neumann said since no one really knows what entropy really is you might as well call it entropy. Shannon's definition applies to discrete random variables X that have values $x_i$. Then if the probability of turning up a given $x_i$ is $p_i$ the Shannon entropy is given by

$$H(X) = -\sum_{i=1}^{i=n} p(x_i) \log(p(x_i)). \qquad (A1)$$

That this had any relation to the Gibbs entropy about which I knew nothing or to anything else I had been reading about entropy, I had not the slightest idea. I just thought that it was quite wonderful whatever it was.

That the statistical mechanical entropy had something to do with information became clearer to me over time. Qualitatively if you double the number of molecules, all things being equal, you will double the entropy and it will require more information to specify the state of the system. It is this general idea that I am going to use to give an intuitive understanding of the Bekenstein-Hawking black hole entropy. I will be following a treatment given by P.C.W.Davies[18] We cannot expect that this treatment will produce the numerical constants. For that you will need a real calculation. In the spirit of this treatment I will set any number of order one to be one. For example pi will be set to one. Moreover I will confine this derivation to the simplest species of black hole. This black hole is uncharged and does not rotate. Here is the idea.

From what I have said the entropy should be proportional to the number of particles that constitute the black hole. Here we immediately run into the basic problem. Viewed from the exterior-beyond the event horizon-we have no idea of what the composition of the black hole is. The pre-Hawking black holes absorbed everything and emitted nothing. They were characterized by three parameters; the mass, the charge and the angular momentum of rotation. This is all an observer outside the event horizon can ever know. The Hawking radiation does not change this general perspective. It is black body radiation and black bodies are notorious from how little they disclose of their inner nature. Nonetheless we want to estimate the number of particles a black hole of mass M is made up of. If we knew the mass m of these particles-assuming they are or one type-that number would be given by M/m. But I am going to ask a somewhat different question. What is the maximum number of such particles? We could get a handle on this if we knew the minimum mass that such particles could have. If the problem were classical there would be no way of knowing and our enterprise would founder. But quantum mechanics saves us.

The radius of our black hole-the Schwarzschild radius is given in terms of the mass M by

$$r = 2GM/c^2 \sim GM/c^2. \qquad (A2)$$

Here M is the mass of the black hole and G is the gravitational constant $6.67 \times 10^{-11}$ $m^{3/}/kgs^2$ . For a particle to fit it must have a deBroglie wave length shorter than this. But that wave length is inversely proportional to the mass that that the minimum mass is given by

$$m \sim hc/GM. \qquad (A3)$$

Thus the number of such particles is given by

$$N \sim M^2 G/hc. \qquad (A4)$$

But by our general notion the entropy should be proportional to this number. But note that the surface area of this black hole is

$$A \sim G^2 M^2/c^4. \qquad (A5)$$

This gives us an understanding of why the Bekenstein-Hawking entropy is proportional to the surface area of the black hole divided by Planck's constant. It tells us that in the classical limit the entropy is infinite.

We see that

$$S \sim M^2 G/hc. \qquad (A6)$$

The change in energy of the black hole comes about when there is a change in mass. Something is dropped into the hole increasing its mass. Thus

$$dMc^2 \sim TdS, \qquad (A7)$$

or

$$dS/dMc^2 \sim 1/T, \qquad (A8)$$

or

$$T \sim hc^3/GM . \qquad (A9)$$

Acknowldegements: I am very grateful to Elihu Abrahams, Steve Adler, Kenneth Ford, Cameron Reedf and Robert Swendsen for critical readings of the manuscript and to John Rosner for scrupulous editorial comments.

---